\documentclass[useAMS,usenatbib]{mn2e}
\usepackage{epsfig}
\usepackage{graphicx}
\input epsf

\newcommand{\xmmn}{{\it XMM-Newton~\/}}
\newcommand{\asca}{{\it ASCA~\/}}
\newcommand{\chandra}{{\it Chandra~\/}}

\newcommand{\rosat}{{\it ROSAT~\/}}

\def\ergcms{{\rm ~erg~cm^{-2}~s^{-1}}}

\def\ergsec{{\rm ~erg~s^{-1}}}

\def\H0{{\rm ~km~s^{-1}~Mpc^{-1}}}

\def\la{\mathrel{\hbox{\rlap{\hbox{\lower4pt\hbox{$\sim$}}}{\raise2pt\hbox{$<$}}}}}
\def\ga{\mathrel{\hbox{\rlap{\hbox{\lower4pt\hbox{$\sim$}}}{\raise2pt\hbox{$>$}}}}}
\def\d25{D$_{25}$}

\def\hii{H{\small II}$~$}

\def\deg{\hbox{$^\circ$ }}
\def\arcm{\hbox{$^\prime$ }}
\def\arcs{\hbox{$^{\prime\prime}$ }}

\title [\xmmn observations of NGC~3256 \& NGC~3310] {\xmmn observations of the starburst merger galaxies NGC~3256 \& NGC~3310}
\author[L. Jenkins et al.]
	{L.P.\ Jenkins$^1$\thanks{E-mail: lej@star.le.ac.uk}, T.P.\ Roberts$^1$, M.J.\ Ward$^1$, A.\ Zezas$^2$\\ 
$^1$ X-ray \& Observational Astronomy Group, Dept. of Physics \& Astronomy, University of Leicester, University Road, Leicester LE1 7RH, U.K.\\
$^2$ Harvard-Smithsonian Center for Astrophysics, 60 Garden Street, Cambridge, MA 02138, USA. \\}

\date{Accepted 2004 May 14; Received in original form 2003 December 18}

\pagerange{\pageref{firstpage}--\pageref{lastpage}}
\pubyear{2003}

\begin{document}

\maketitle

\label{firstpage}

\begin{abstract}

We present \xmmn EPIC observations of the two nearby starburst merger galaxies NGC~3256 \& NGC~3310. The broad-band (0.3--10\,keV) integrated X-ray emission from both galaxies shows evidence of multi-phase thermal plasmas plus an underlying hard non-thermal power-law continuum. NGC~3256 is well-fit with a model comprising two MEKAL components ($kT$=0.6/0.9\,keV) plus a hard power-law ($\Gamma$=2), while NGC~3310 has cooler MEKAL components ($kT$=0.3/0.6\,keV) and a harder power-law tail ($\Gamma$=1.8). \chandra observations of these galaxies both reveal the presence of numerous discrete sources embedded in the diffuse emission, which dominate the emission above $\sim$2\,keV and are likely to be the source of the power-law emission. The thermal components show a trend of increasing absorption with higher temperature, suggesting that the hottest plasmas arise from supernova-heated gas within the disks of the galaxies, while the cooler components arise from outflowing galactic winds interacting with the ambient interstellar medium (ISM). We find no strong evidence for an active galactic nucleus (AGN) in either galaxy.

\end{abstract}


\section{Introduction}
\label{sec:intro}
By virtue of their greatly enhanced star formation rates, starburst galaxies (SBGs) are very powerful sources of X-ray emission, and consequently exhibit complex spectral characteristics reflecting the variety of energetic phenomena related to the end points of stellar evolution. Starburst episodes are transitory events in the lifetimes of galaxies, with enhanced SFRs that are unsustainable over periods greater than $\sim10^8$\,yr (e.g. \citealt{heckman98}). Galaxies going through starburst phases appear to be a common phenomena in the early Universe (e.g. \citealt*{smail97}), and local SBGs show characteristics similar to high-redshift examples. Therefore, our understanding of the X-ray emission processes in local systems is of fundamental importance. For example, SBGs are believed to be the prime source of ``stellar feedback'', whereby supernova-driven outflows (winds) enrich both the galaxy's interstellar medium (ISM) and the intergalactic medium (IGM)\citep{strickland00b}. The similarity of local and high-redshift examples is further highlighted by the detection of large-scale galactic winds in UV observations of star-forming Lyman break galaxies at z $\simeq$ 3 (\citealt{pettini00}; \citealt{pettini01}). 

Mergers of galaxies appear to trigger the most extreme starburst episodes. Merger systems tend to have extremely high infrared (IR) luminosities (luminous IR galaxies (LIRG): $L_{IR}>10^{11}L_\odot$ and ultraluminous IR galaxies (ULIRG): $L_{IR}>10^{12}L_\odot$, e.g. \citealt{sanders96}), the brightest of which can even overlap the bolometric luminosities of quasars and Seyfert galaxies (\citealt{lipari00}; \citealt{lira02} and references therein).  Many luminous IR galaxies are detected in deep-IR surveys, showing that they are an important constituent of the high-redshift ($z\ga1$) Universe (e.g. \citealt{franceschini01}). In the hierarchical merger hypothesis (e.g. \citealt{toomre77}), collisions and mergers of galaxies are thought to be one of the most dominant evolutionary mechanisms for galaxies, whereby isolated disk galaxies merge through close tidal encounters to eventually form elliptical galaxies (\citealt{kormendy92}; \citealt{read98}; \citealt{genzel01}).  X-ray observations of nearby systems with obvious signatures of merging at other wavelengths provide a window through which to search for the high-energy signatures of their primary sources of power i.e. starburst activity or the presence of an active galactic nucleus (AGN).

Prior to the \xmmn and \chandra era, X-ray mission such as {\it Einstein}, \rosat and \asca revealed the signatures of the various processes present in the closest local SBGs. For example, \rosat X-ray studies of the soft (0.1--2.4\,keV) emission in the two archetypal SBGs M82 and NGC~253 showed that both had substantial extended diffuse emission from the disks and halo regions that were well modelled with thermal plasmas with $kT\la1$\,keV (\citealt*{strickland97}; \citealt{pietsch00}), whereas broader band (0.5--10\,keV) \asca studies of these and other SBGs showed that their spectra could be described by models comprising one- or two-temperature thermal plasma components ($kT\la1$\,keV) plus a harder component fitted with either high-temperature plasma or a non-thermal power-law (e.g., M82 \& NGC~253, \citealt{ptak97};  NGC~3310 \& NGC~3690, \citealt*{zezas98}; NGC~3256, \citealt*{moran99}).

However, subsequent to the launch of the \xmmn and \chandra X-ray observatories, we are now able to utilize their excellent complimentary spectral and imaging capabilites to study the detailed spatial and spectral characteristics of these X-ray emission processes. The sub-arcsecond spatial resolution of \chandra can resolve individual compact X-ray sources and accurately determine their positions, whereas the high throughput of \xmmn can yield detailed spectral information of bright point sources and low-surface-brightness extended emission. For example, \xmmn observations have given an unprecedented wide-field view of the spectral properties of the diffuse emission in the nucleus, disk and halo in NGC~253 \citep{pietsch01}, while \chandra observations gave detailed images of point sources and diffuse emission in the central regions of the galaxy (\citealt{strickland00b}; \citealt{strickland02}). Similar work has also been done with M82 (\citealt*{stevens03}; \citealt{matsumoto01}).

The main X-ray spectral components of SBGs have recently been quantified by \citet{persic02}, based on the observed spectral properties of M82 \& NGC~253 as derived from {\it BeppoSAX} observations and a stellar-population evolutionary model. They assessed contributions from the gaseous and stellar X-ray processes expected in SBGs, i.e. from X-ray binaries (XRBs), supernova remants (SNRs), Compton scattering of ambient far-infrared (FIR) photons off supernova-accelerated relativistic electrons, diffuse thermal plasma and a compact nucleus in the form of a starburst or AGN. Their results show that at low energies ($\la$ 2\,keV), the {\it dominant} component is expected to be a low temperature ($kT\leq1$\,keV) diffuse plasma resulting in part from the galactic wind itself, but mainly from shock heating via the interaction of the hot low-density wind with the ambient high-density ISM \citep{strickland00a}. At higher energies (2--10\,keV), they predict that the spectrum will be {\it dominated} by emission from bright neutron star XRBs (high- and low-mass secondaries) in the form of a cut-off power-law with a possible contribution from non-thermal Compton emission or an AGN if present \citep{persic02}.

However, recent studies of SBGs with \xmmn and {\it Chandra} have shown that, in the absence of a bright AGN, the X-ray output of starforming galaxies are dominated (particularly in the hard 2--10\,keV band) by small numbers of bright extra-nuclear point sources (e.g. \citealt{kilgard02}; \citealt{colbert04}), which show increasing evidence of being XRB systems, possessing black hole rather than neutron star primaries (e.g. \citealt{fabbianowhite04}; \citealt{miller04}). The brightest of these discrete X-ray sources are the ultraluminous X-ray sources (ULXs), defined as those possessing X-ray luminosities $> 10^{39} \ergsec$, which exceeds the Eddington limit for accretion onto a 1.4$M_{\odot}$ neutron star (see section~\ref{sec:discuss_hard} for further discussion).

In this paper, we present \xmmn X-ray observations of the two nearby starburst merger systems NGC~3256 \& NGC~3310, together with complementary published \chandra data for NGC~3256 and new \chandra results for NGC~3310 (Zezas et al. {\it in preparation}) to aid our interpretation of the spectral information from {\it XMM-Newton}. This paper is set out as follows. In sections~\ref{sec:intro_3256} and~\ref{sec:intro_3310}, we give a summary of the known properties of each galaxy based on multiwavelength studies. In section~\ref{sec:data}, we outline the \xmmn data reduction methods used, then in section~\ref{sec:results} we present the detailed results of the spectral fitting procedures and compare these with previous results from {\it ROSAT}, \asca and \chandra observations. We discuss our results in section~\ref{sec:discuss} in the context of other recent results from studies with \xmmn and {\it Chandra}, and discuss how these compare as a whole to the starburst spectral template of \citet{persic02}. Our conclusions are summarised in section~\ref{sec:conclusions}. The distances we adopt in this paper are taken from \citet{sanders03}, which assume a value of $H_0$=75\,km\,s$^{-1}$\,Mpc$^{-1}$.

\section{The Galaxies}

\begin{table*}
\caption{Details of the \xmmn observations.}
 \centering
  \begin{tabular}{lcccccccccc}
\hline
Target    & Observation ID & Date         & & \multicolumn{2}{c}{Useful exposure (s)} & & Filter & & \multicolumn{2}{c}{Source count rate (ct s$^{-1}$, 0.3--10\,keV)}\\
          &                & (yyyy-mm-dd) & & PN            & MOS                     & &        & & PN           & MOS            \\ 
\hline
NGC~3256  & 0112810201     & 2001-12-15   & & 6862          & 11370                   & & Thin   & & 0.55         & 0.16            \\
NGC~3310  & 0112810301     & 2001-05-11   & & 6728          & 8764                    & & Thin   & & 0.67         & 0.19            \\
\hline
\end{tabular}
\label{table:obs}
\end{table*}

\subsection{NGC~3256}
\label{sec:intro_3256}

NGC~3256 is a starburst merger system located at a distance of D=35.4\,Mpc. It is IR-bright ($L_{FIR}=2.7\times10^{11}L_\odot$, \citealt{sanders03}), although it does not reach the ULIRG category.  Its starburst nature has been confirmed at UV \citep{kinney93}, optical \citep{lipari00}, IR (\citealt{graham84}; \citealt{doyon94}; \citealt{moorwood94}) and radio wavelengths \citep{norris95}. It shows a highly disturbed structure with prominent extended tidal tails spanning $\sim$80\,kpc, which are believed to be the signature of a merger between two gas-rich galaxies of roughly equal size (\citealt{toomre72}; \citealt{english03}). This merger is thought to be relatively young ($5\times10^8$ yr, \citealt{lipari00}), because within the highly disturbed central region two distinct nuclei have been resolved at radio \citep{norris95}, IR \citep{kotilainen96} and optical \citep{lipari00} wavelengths.  The northern nucleus has been identified as a pure starburst with a powerful ouflowing galactic wind \citep{lipari00}. Despite previous speculation based on near-IR and radio observations that the southern nucleus may host an AGN (e.g. \citealt{kotilainen96}; \citealt{norris95}), this seems increasingly unlikely as no unambiguous evidence for AGN activity was found in the recent \chandra X-ray observation which resolves both nuclei \citep{lira02}, although a recent radio study by \citet*{neff03} detected strong radio emission from both nuclear sources as well an ULX and show that both nuclei have properties consistent with radio-loud LLAGN. The age of the starburst component in this galaxy is $\sim5-25\times10^{6}$ yr, which means that it must have started long after the merger began \citep{lipari00}. This galaxy was also observed with earlier X-ray missions, i.e. \rosat \citep{boller92} and \asca \citep{moran99}, and has been found to have a high X-ray luminosity (observed $L_X\sim2\times10^{41} \ergsec$ in the 0.5-10\,keV band, \citealt{moran99}).

\subsection{NGC~3310}
\label{sec:intro_3310}

NGC~3310 is a closer but smaller system than NGC~3256, located at a distance of $\sim19.8$\,Mpc and classified as type SAB(r)bc pec \citep{devaucouleurs91}. It has a disturbed structure, and the starburst (aged $\sim10^7-10^8$ yrs, \citealt{elmegreen02} and references therein) is thought to have been triggered by a collision between NGC~3310 and a dwarf companion during the last $\sim10^8$\,yr \citep{balick81}. Although no double nucleus has been resolved in NGC~3310, further evidence of a merger comes from kinematic peculiarities in that the centre of rotation of the emission line gas does not coincide with its nucleus (\citealt{vanderkruit76}; \citealt{balick81}).  One of the most striking features of the optical emission is a prominent ``bow \& arrow'' structure extending 100 arcseconds northwest of the galaxy's centre, which is believed to be a remnant of the merger \citep{balick81}. Its FIR luminosity ($L_{FIR}=2.8\times10^{10}L_\odot$, \citealt{sanders03}) is comparable to that of M82, although a factor of ten less luminous than NGC~3256. NGC~3310 has been studied in numerous multiwavelength observations, including optical (\citealt{balick81}; \citealt{pastoriza93}; \citealt{mulder96}), ultraviolet (UV) (\citealt{vanderkruit76}; \citealt{meurer95}; \citealt{smith96}), IR (\citealt{telesco84}; \citealt{pastoriza93}), H$\alpha$ (\citealt{mulder96}; \citealt{conselice00}) and radio (\citealt{balick81}; \citealt{vanderkruit76}; \citealt*{mulder95}; \citealt{kregel01}). Early optical and H$\alpha$ images showed that the bright inner regions are dominated by an open spiral arm pattern (e.g. \citealt{balick81}), and the inner part of this region connects to a $\sim30$ arcsecond ($\sim2.9$\,kpc) diameter starburst ring, which in turn surrounds a blue compact nucleus \citep{kregel01}.  Optical and near-IR spectrophotometric observations have shown that the nucleus has solar abundances, whereas the circumnuclear and disk \hii regions show a lower metallicity \citep{pastoriza93}.  There is a ``jumbo'' \hii region $\sim12$ arcseconds southwest of the nucleus, whose size and H$\alpha$ luminosity is comparable to the largest extragalactic \hii regions known (e.g. NGC~5471 in M101, \citealt{balick81}).  Most recently, high-resolution {\it Hubble Space Telescope} (HST) optical (WFPC2) and IR (NICMOS) observations of NGC~3310 show the presence of numerous super star clusters in the innermost southern spiral arm and circumnuclear ring whose positions correlate with radio and H$\alpha$ peaks, with ages consistent with their formation resulting from the cannibalism of a dwarf galaxy in the last 10\,Myr \citep{elmegreen02}. It is important to note that there is no evidence of AGN activity in this system at any wavelength. For example, no broad H$\alpha$ line is detected by \citet*{ho97}, which would be expected from an AGN with the hard (2--10\,keV) X-ray luminosity measured from \rosat and \asca observations assuming the $L_X/L_{H\alpha}$ relation of \citet*{elvis84} (c.f. \citealt{zezas98}).

\begin{figure*}
\centering
\scalebox{0.68}{\includegraphics{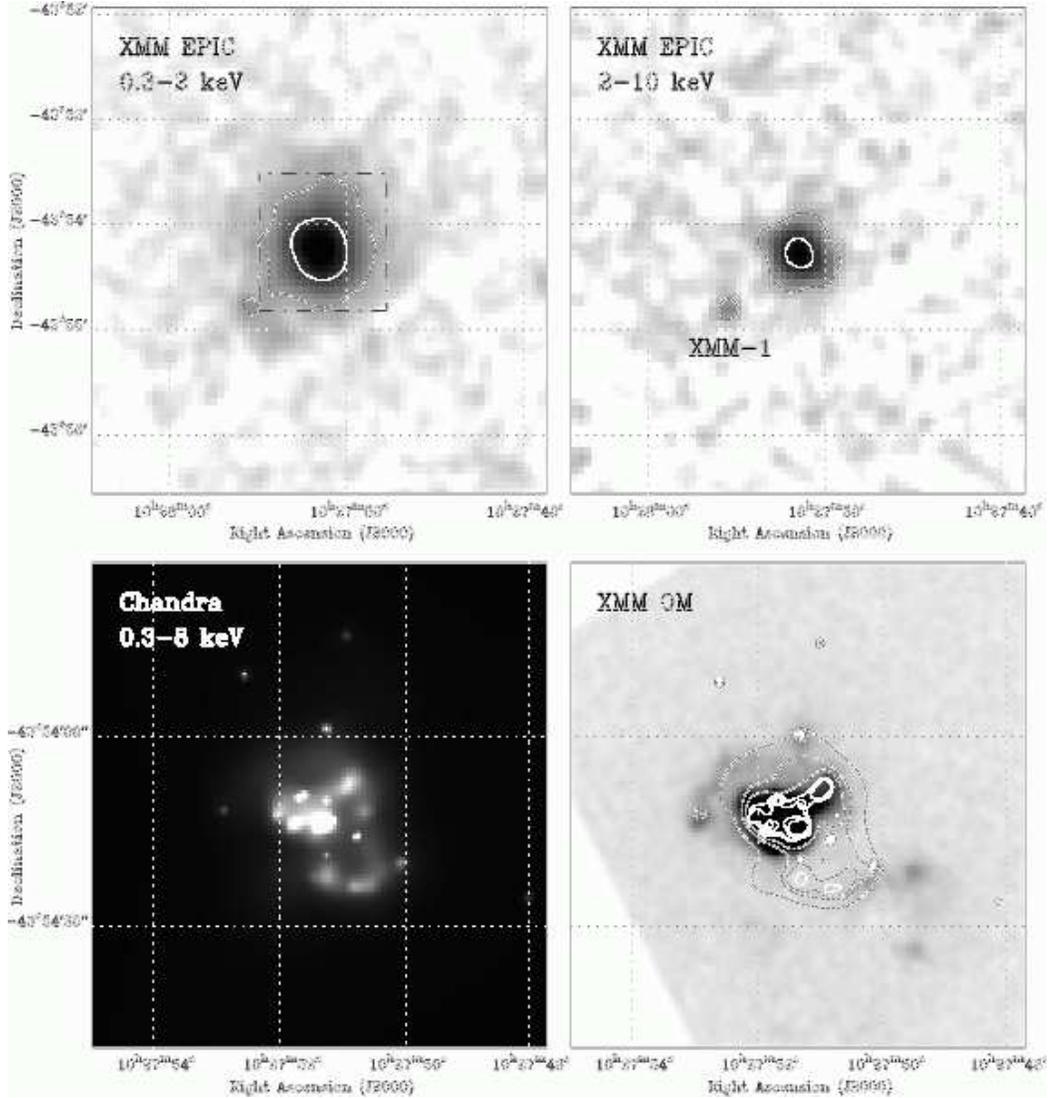}}
\caption{[Top] \xmmn EPIC soft (0.3--2\,keV, left) \& hard (2--10\,keV, right) X-ray images of NGC~3256. The images are the sum of all 3 detectors, and are convolved with a 1 pixel (4 arcsecond) HWHM 2-D Gaussian mask and displayed with a logarithmic scale. The intensity minima/maxima correspond to surface brightnesses of $3\times10^{-5}$/$3\times10^{-3}$ ct s$^{-1}$ (soft) and $3\times10^{-5}$/$8\times10^{-4}$ ct s$^{-1}$ (hard), and the contours correspond to $2\times10^{-4}$/ $1\times10^{-3}$ ct s$^{-1}$ (soft) and $8\times10^{-5}$/$5\times10^{-4}$ ct s$^{-1}$ (hard). [Lower left] {\it Chandra} (0.3--8\,keV) linearly scaled image of the (dot-dashed) boxed region shown in the \xmmn image, adaptively smoothed with a 3$\sigma$ local significance threshold. [Lower right] {\it Chandra} X-ray contours corresponding to surface brightnesses of [0.4,0.7,1.1,1.8,2.1,3.2,4.3]$\times10^{-4}$ ct s$^{-1}$ overlaid on an \xmmn OM UVW2 ($\lambda_{central}$=2120\AA) image of NGC~3256.}
\label{fig:n3256}
\end{figure*}

\section{Observations, Data Reduction \& Analysis Techniques}
\label{sec:data}

\subsection{\xmmn Observations \& Reduction}

The details of the \xmmn observations are shown in Table~\ref{table:obs}. NGC~3256 was observed for 17\,ks on the 15th Dec, 2001; NGC~3310 for 19\,ks on the 11th May, 2001.  During both observations the EPIC MOS-1, MOS-2 \& PN cameras were operated with thin filters in ``Prime Full Window'' mode. The event lists were pipeline-processed using the {\small SAS} ({\it Science Analysis Software}) {\small v5.2.1} (NGC~3256) and {\small v5.1.0} (NGC~3310) and all data products (spectra, images \& lightcurves) were created using {\small SAS} {\small v5.4.1} and the EPIC calibration as of March 2003. Full-field light curves were accumulated for all exposures in the two observations to check for high background intervals of soft proton flares. In the case of NGC~3256, there were numerous small flares throughout the observation and a large section of high level flaring for the last $\sim$3\,ks. Time intervals with count rates $>$ 50\,ct s$^{-1}$ (PN) and 15\,ct s$^{-1}$ (MOS) were cut from subsequent data analysis, leaving a net good time for each camera of $\sim$11\,ks (MOS 1 \& MOS 2) and $\sim$7\,ks (PN). For NGC~3310, there was a high level of flaring throughout almost the entire observation. In order to extract any useful information from the data, we accepted time intervals with relatively high background count rates ($<$ 140\,ct s$^{-1}$ [PN] \& $<$ 30\,ct s$^{-1}$ [MOS]), providing us with relatively clean data below 6\,keV. This left net good times of $\sim$9\,ks (MOS 1 \& MOS 2) and $\sim$7\,ks (PN).

\begin{table*}
\caption{Spectral fitting results for NGC~3256.}
 \centering
  \begin{tabular}{cccccccccccc}        
\hline
\multicolumn{2}{c}{PL} & & \multicolumn{2}{c}{M$_1$} & & \multicolumn{2}{c}{M$_2$}   &  $\chi^2$/dof  & $F_X$$^b$     & \multicolumn{2}{c}{$L_X$$^c$}   \\ 

\\[-3mm]
$N_H$$^a$              & $\Gamma$  & & $N_H$$^a$ & $kT (keV)$    & & $N_H$$^a$  & $kT (keV)$  & &              &  Obs            & Unabs         \\
\hline
\\

\multicolumn{12}{l}{{\bf Model 1:} M+PL ({\it wabs*po+wabs*mekal})}\\
\\
3.06$^{+0.25}_{-0.24}$ & 2.70$^{+0.13}_{-0.13}$ & & $\dag$ & 0.64$^{+0.02}_{-0.02}$ & & - & - &  387.1/300 & 9.71$^{+0.45}_{-0.56}$ & 1.46$^{+0.07}_{-0.08}$ & 3.78$^{+0.10}_{-0.44}$ \\
\\

\multicolumn{12}{l}{{\bf Model 2:} M+M+PL* ({\it wabs*po+wabs*mekal+wabs*mekal})}\\
\\
7.66$^{+1.08}_{-0.95}$                & 2.31$^{+0.27}_{-0.48}$ & & Gal  & 0.35$^{+0.04}_{-0.02}$ & & $\dag$  & 0.73$^{+0.05}_{-0.04}$ &  355.7/298  & 9.91$^{+0.42}_{-1.08}$  & 1.49$^{+0.06}_{-0.16}$ & 5.74$^{+0.06}_{-0.92}$ \\
\\

\multicolumn{12}{l}{{\bf Model 3: M+M+PL* {\it (wabs*po+wabs*mekal+wabs*mekal)}}}\\
\\
{\bf 1.47$^{+1.05}_{-0.44}$} & {\bf 1.98$^{+0.41}_{-0.40}$} & & {\bf Gal} & {\bf 0.57$^{+0.05}_{-0.11}$} & & {\bf 9.47$^{+1.53}_{-1.24}$} & {\bf 0.85$^{+0.08}_{-0.10}$} & {\bf 343.3/297}  & {\bf 10.00$^{+0.53}_{-1.59}$} & {\bf 1.50$^{+0.08}_{-0.24}$} & {\bf 4.53$^{+0.26}_{-1.17}$} \\
\\

\hline
\end{tabular}
\begin{tabular}{l}
Notes: Spectral models: PL=power-law continuum model and M=MEKAL thermal plasma (solar abundances). ~$^a$Absorption \\column in units of $10^{21}$ cm$^{-2}$. ~$^b${\it Observed} fluxes in the 0.3--10\,keV band, in units of $10^{-13}$ erg s$^{-1}$ cm$^{-2}$. ~$^c${\it Observed} and \\{\it unabsorbed} luminosities in the 0.3--10\,keV band, in units of $10^{41} \ergsec$ (assuming a distance of 35.4\,Mpc). ~$\dag$Same hydrogen column \\as applied to the power-law spectral component. ~$^*$Models with the cool MEKAL $N_H$ component fixed at the Galactic value. Model \\parameter errors correspond to 90 percent confidence limits for 1 parameter of interest. The best-fit model is shown in bold.\\
\end{tabular}
\label{table:n3256}
\end{table*}

\subsection{\xmmn Data Analysis}
\label{sec:analysis}

The X-ray spectra of NGC~3256 \& NGC~3310 were extracted using circular regions of radii 45 and 40 arcseconds respectively. The background regions were taken close to each source, using as large an area as possible with approximately the same DET-Y distance as the source region (to ensure similar low-energy noise subtraction).  We used the {\small SAS} task {\small ESPECGET} with standard filtering to simultaniously extract source and background spectra as well as create response matrices (RMFs) and ancillary response files (ARFs) for each source. The resulting spectra were binned to a minimum of 20 counts per bin in order to optimise the data for $\chi^2$ statistics.

The spectral analysis of the \xmmn data in this study has been performed using {\small XSPEC 11.1/11.3}. All errors are given at the 90 percent confidence level unless stated otherwise.  For NGC~3256, the MOS and PN spectra have been fitted simultaneously in the 0.3-10\,keV band, whereas we have only fitted the NGC~3310 data in the 0.3--6\,keV range as the high-energy data were heavily contaminated by soft proton flaring. We have also included a free normalisation constant in the models to account for differences in the flux calibration of the three EPIC cameras, which differ by $\la$ 15 percent in practise.

Following previous work in this field, the two spectral models we use to fit the data are a power-law continuum to represent the combined emission from compact sources, and an optically-thin thermal plasma (MEKAL) to model the thermal components expected in SBGs such as galactic winds and SNRs. For this study we have chosen to fix the abundance parameter in the MEKAL models to solar values, as previous results with \rosat and \asca data have shown that fitting complicated multi-temperature thermal plasmas with this type of simple spectral model tends to result in unrealistically low metal abundances of $\sim$0.05--0.3Z${\odot}$ (\citealt{strickland00a} and references therein).

\section{Results}
\label{sec:results}

\subsection{NGC~3256}
\label{sec:n3256_results}

In Figure~\ref{fig:n3256} we show \xmmn EPIC images of NGC~3256 in soft (0.3--2\,keV, top left) and hard (2--10\,keV, top right) bands, plus an adaptively smoothed archival \chandra image of the central region [lower left]. No X-ray structure is resolved in the main body of the galaxy in the \xmmn data, but one discrete source is detected to the southeast (labelled XMM-1, see section~\ref{sec:n3256_discrete}). However, the \chandra images do resolve the bulk of the emission into several discrete sources embedded within diffuse emission. The results of the \chandra analysis are presented in \citet{lira02}, where it is shown that the integrated emission from the galaxy is best fit with a model comprising two MEKAL components plus a power-law continuum. This study shows that the hard power-law component with an index of $\Gamma\sim2$ is mainly (75--80 percent) due to the fourteen compact discrete sources (including the two galactic nuclei), which contribute $\ga$ 20 percent of the total emission of the galaxy in the 0.5--10\,keV range. \citet{lira02} speculate that the remainder of the hard emission is likely to come from unresolved XRBs and supernovae distributed throughout the galaxy, and that one of the thermal components (0.6\,keV) is the signature of the hot superwind, while the other (0.9\,keV) arises from SNR heated gas within a much smaller volume near the starburst nucleus. Figure~\ref{fig:n3256} [lower right] shows the \chandra X-ray contours overlaid on an \xmmn Optical Monitor (OM) UV image of NGC~3256 of the same scale. This illustrates that the UV emission, which typically arises in starforming regions, correlates well with the \chandra X-ray data, although a section of the UV emission is obscured by the presence a large dust lane in the southwest of the galaxy (see Figure~10 in \citealt{lira02}).

\begin{figure}
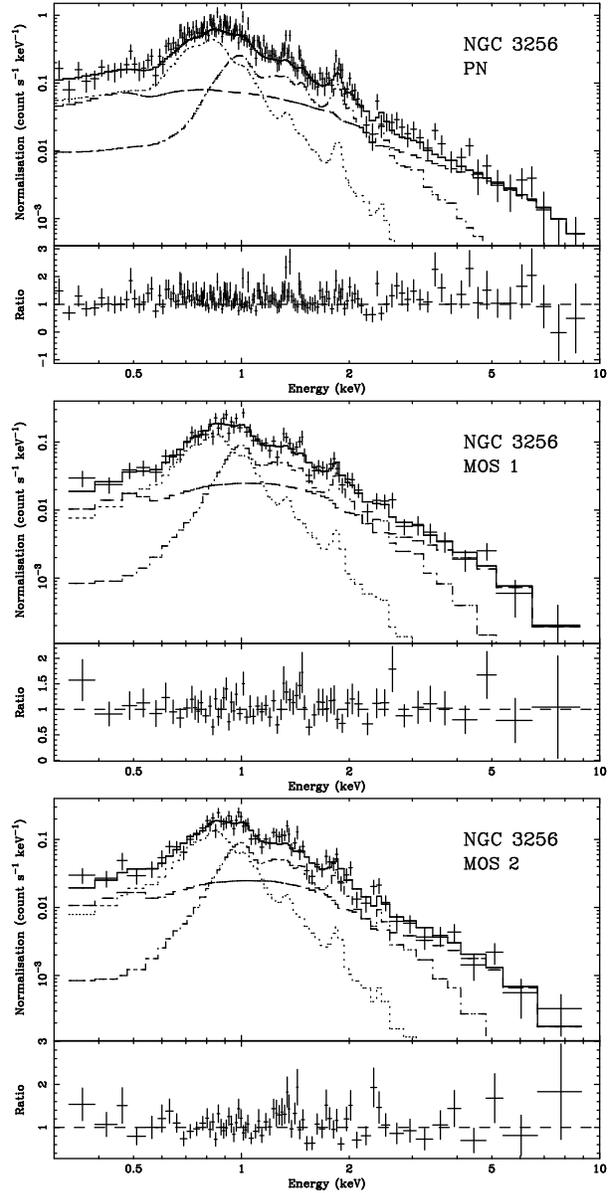

\centering
\scalebox{0.34}{\rotatebox{270}{\includegraphics{figure2a.ps}}}
\scalebox{0.34}{\rotatebox{270}{\includegraphics{figure2b.ps}}}
\scalebox{0.34}{\rotatebox{270}{\includegraphics{figure2c.ps}}}
\caption{\xmmn spectra and folded best-fit model (M+M+PL) of NGC~3256: PN [top], MOS 1 [middle], MOS 2 [bottom]. The model components are denoted by dashed (power-law), dotted (cool MEKAL) \& dot--dashed (warm MEKAL) lines.}
\label{fig:3256spec}
\end{figure}

Since we do not spatially resolve any X-ray structure in the centre of NGC~3256 with {\it XMM}, we can only fit the spectrum of the integrated emission. We began by fitting the data with single- and two-component spectral models with combinations of power-law and MEKAL components, but we rejected these due to unacceptable values of chi-squared ($\chi^2_{\nu}>2$), although a M+PL model with equal absorbing columns did yield a reasonable fit as shown in Table~\ref{table:n3256}. In order to improve this, we went on to fit the data with three-component models (M+M+PL) to better represent the multi-temperature gas expected in this system. We also included separate absorbing $N_H$ columns for each of the three components, but fixed the absorption in the coolest MEKAL component to the Galactic value (9.5$\times10^{20}$ cm$^{-2}$, \citealt{dickey90}) as fitting this parameter resulted in $N_H$ values below the Galactic value.  Following the work of \citet{lira02}, we tried a model where both absorbing $N_H$ components were fitted independently, and one where they were tied together. The results are shown in Table~\ref{table:n3256}: the best-fit model (shown in bold) is the one with independent absorbing columns with $kT$=0.57/0.85\,keV, $\Gamma$=1.98 and $\chi^2_{\nu}=1.16$, which is a $>$ 99.9 percent improvement over the M+PL model according to the F-test statistic. The PN and MOS spectra are shown with the best-fit model in Figure~\ref{fig:3256spec}, and Table~\ref{table:n3256_comps} lists the observed fluxes, unabsorbed luminosities and overall percentage contributions to the total of each from the three spectral components in the \xmmn fit.

The errors on the fit parameters in Table~\ref{table:n3256} correspond to 90\% confidence limits for 1 parameter of interest ($\Delta\chi^2$=2.71). However, the temperatures and absorption values for multiple thermal components in models such as these are strongly correlated. The absorption column associated with the cool thermal component is fixed in the best-fit model, but more realistic errors for the warm component i.e. for 2 parameters of interest for a 90\% confidence limit ($\Delta\chi^2$=4.61) are practically unchanged with $N_H=8.0-11.4\times10^{21}$ cm$^{-2}$ and $kT$=0.74--0.95\,keV, demonstrating that these parameters are well constrained.

The PN data also show possible line emission at $\sim$6--7\,keV, which could be a result of either neutral 6.4\,keV Fe-K emission from an AGN or helium-like Fe 6.7\,keV emission from a population of type Ib/IIa SNRs residing in regions of star-formation (e.g. \citealt{behar01}). We have therefore fitted a narrow gaussian line to the power-law continuum of the best-fit model, which results in an unconstrained line energy of $\sim$ 6.5\,keV with an equivalent width of 250\,eV. Although the addition of the gaussian component only improves the fit statistics at the 1$\sigma$ level, the 90 percent upper limit of the line normalisation allows us to constrain an upper limit of 1045\,eV for the equivalent width of the line. We will discuss the implications of this possible line emission in section~\ref{sec:discuss}.

\begin{table}
\caption{Fluxes, luminosities and percentage of total emission from the three components of the best-fit M+M+PL model of NGC~3256.}
 \centering
  \begin{tabular}{lccccc}
\hline
Component  &$F_X$  &\% of total & $L_X$ & \% of total \\
\hline
Cool MEKAL & 2.50  & 25         & 0.53  & 12          \\
Warm MEKAL & 3.35  & 33         & 3.16  & 70          \\
Power-Law   & 4.15  & 42         & 0.84  & 18          \\
\hline
Total      & 10.00 & 100        & 4.53 & 100          \\
\hline
\end{tabular}
\begin{tabular*}{7.5cm}{@{}l}
Notes:  {\it Observed} fluxes in the 0.3--10\,keV band, in units of \\$10^{-13}$ erg s$^{-1}$ cm$^{-2}$. {\it Unabsorbed} luminosities in the \\0.3--10\,keV band, in units of $10^{41} \ergsec$ (assuming a \\ distance of 35.4\,Mpc).\\
\end{tabular*}
\label{table:n3256_comps}
\end{table}

\subsubsection{Comparison with previous work}

Reassuringly, the \xmmn results are remarkably similar to the results of the 30\,ks \chandra observation reported by \citet{lira02}, where the best-fit model for the integrated emission from NGC~3256 is also a M+M+PL model with $kT$=0.58/0.92\,keV and $\Gamma$=1.99. In addition, the absorbing hydrogen columns are consistent between the two datasets within the 90 percent confidence limits. However, the observed flux of the \chandra model in the 0.5--10\,keV range ($\sim1.2\times10^{-12} \ergcms$) is $\sim$20 percent higher than the equivalent \xmmn model in the same energy range ($\sim1.0\times10^{-12} \ergcms$). In order to try to understand this difference, we have compared the relative contributions from the three model components in the \xmmn data (Table~\ref{table:n3256_comps}) with those found in the \chandra data based on the model parameters and component normalisations quoted in Table~7 of \citet{lira02}. This comparison shows that, while the summed contributions from the two thermal components agree to within $\sim$10 percent, there is a much lower contribution from the hard power-law component in the \xmmn observation ($\sim$65 percent of the \chandra flux). This is not an aperture affect, as the \xmmn and \chandra extraction regions are virtually the same size (45 and 40 arcsecond radii respectively). It is much more likely that the difference in the power-law flux is a real variation in the summed flux of the ULX population. If the brightest compact X-ray source (the northern nucleus, see Table~3 in \citealt{lira02}) were to ``switch off'', this would result in a $\sim$21 percent drop in the power-law flux, which would explain the majority of this difference. The alternative is a situation in which there is a coincidental drop from several sources, although this is much less likely.

The results of the fits to the \asca data for NGC~3256 reported by \citet{moran99} showed similar results and model parameters, with the best-fit model comprising two Raymond-Smith (RS) plasma components plus a hard power-law ($kT$=0.29/0.80\,keV, $\Gamma$=1.68). Even though the relative contributions of the \asca model components differ from both the \chandra and \xmmn results, the overall observed flux in the 0.5--10\,keV range ($\sim1.3\times10^{-12} \ergcms$) agrees with the observed \chandra flux to within $\sim$5 percent \citep{lira02}, though it is $\sim$30 percent higher than the \xmmn flux.

\subsubsection{Discrete Source Properties}
\label{sec:n3256_discrete}

\begin{figure*}
\centering
\scalebox{0.68}{\includegraphics{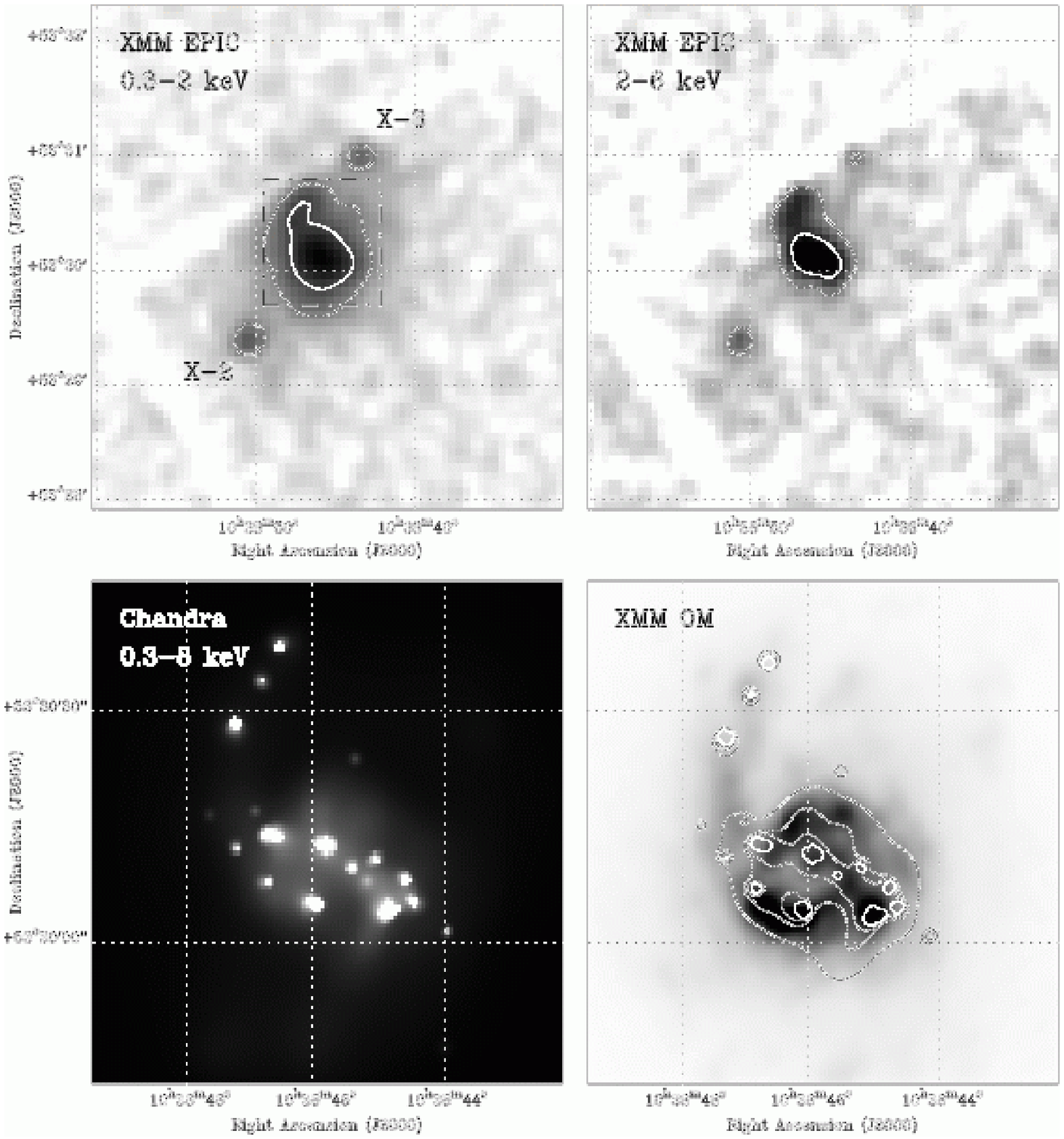}}
\caption{[Top] \xmmn EPIC soft (0.3--2\,keV, left) \& hard (2--6\,keV, right) X-ray images of NGC~3310. The images are the sum of all 3 detectors, and are convolved with a 1 pixel (4 arcsecond) HWHM 2-D Gaussian mask and displayed with a logarithmic scale. The intensity minima/maxima correspond to surface brightnesses of $5\times10^{-5}$/$4\times10^{-3}$ ct s$^{-1}$ (soft) and $9\times10^{-5}$/$8\times10^{-4}$ ct s$^{-1}$ (hard), and the contours correspond to $4\times10^{-4}$/ $1\times10^{-3}$ ct s$^{-1}$ (soft) and $3\times10^{-4}$/$6\times10^{-4}$ ct s$^{-1}$ (hard). [Lower left] {\it Chandra} (0.3--8\,keV) linearly scaled image of the (dot-dashed) boxed region shown in the \xmmn image, adaptively smoothed with a 3$\sigma$ local significance threshold. [Lower right] {\it Chandra} X-ray contours corresponding to surface brightnesses of [2,4,6,16]$\times10^{-5}$ ct s$^{-1}$ overlaid on an \xmmn OM UVW1 ($\lambda_{central}$=2910\AA) image of NGC~3310.}
\label{fig:n3310}
\end{figure*}

\begin{table*}
\caption{Spectral fitting results for NGC~3310.}
 \centering
  \begin{tabular}{cccccccccccc}        
\hline
\multicolumn{2}{c}{PL} & & \multicolumn{2}{c}{M$_1$} & & \multicolumn{2}{c}{M$_2$}   &  $\chi^2$/dof  & $F_X$$^b$     & \multicolumn{2}{c}{$L_X$$^c$}   \\ 
$N_H$$^a$              & $\Gamma$  & & $N_H$$^a$ & $kT (keV)$    & & $N_H$$^a$  & $kT (keV)$  & &              &  Obs            & Unabs         \\

\hline
\\

\multicolumn{12}{l}{{\bf Model 1:} M+PL ({\it wabs*po+wabs*mekal})}\\
\\
1.19$^{\dag}$   & 1.70$^{+0.07}_{-0.08}$ & & 2.97$^{+1.21}_{-0.78}$ & 0.26$^{+0.03}_{-0.03}$ & & -             & -      &   322.8/348 & 17.16$^{+0.67}_{-1.84}$ & 0.81$^{+0.03}_{-0.09}$ & 1.40$^{+0.06}_{-0.47}$ \\
\\

\multicolumn{12}{l}{{\bf Model 2:} M+M+PL ({\it wabs*po+wabs*mekal+wabs*mekal})}\\
\\
1.19$^{\dag}$  & 1.56$^{+0.10}_{-0.07}$ & & 0.42$^{+4.10}_{-0.39}$  & 0.24$^{+0.04}_{-0.09}$ & & 0.86$^{+2.00}_{-0.45}$ & 0.63$^{+0.07}_{-0.05}$ & 288.2/345 & 18.13$^{+0.30}_{-2.05}$ & 0.85$^{+0.01}_{-0.10}$ & 0.99$^{+0.01}_{-0.09}$ \\
\\

\multicolumn{12}{l}{{\bf Model 3:} M+PL ({\it wabs*pcfabs*po+wabs*mekal})}\\
\\
1/10$^{\ddag}$ & 1.98$^{+0.07}_{-0.07}$ & & 4.17$^{+0.74}_{-0.83}$ & 0.24$^{+0.03}_{-0.02}$ & & -             & -       &   309.6/348 & 16.53$^{+0.70}_{-2.07}$ & 0.78$^{+0.03}_{-0.10}$ & 2.21$^{+0.11}_{-0.89}$ \\
\\

\multicolumn{12}{l}{{\bf Model 4: M+M+PL ({\it wabs*pcfabs*po+wabs*mekal+wabs*mekal})}}\\
\\
{\bf 1/10$^{\ddag}$} & {\bf 1.84$^{+0.11}_{-0.13}$} & & {\bf 0.61$^{+1.84}_{-0.13}$} & {\bf 0.25$^{+0.04}_{-0.08}$} & & {\bf 2.19$^{+2.08}_{-1.06}$} & {\bf 0.62$^{+0.06}_{-0.05}$} & {\bf 283.1/345} & {\bf 17.30$^{+0.46}_{-1.82}$} & {\bf 0.81$^{+0.02}_{-0.09}$} & {\bf 1.26$^{+0.01}_{-0.18}$} \\
\\

\hline
\end{tabular}
\begin{tabular}{l}
Notes: Spectral models: PL=power-law continuum model and M=MEKAL thermal plasma (solar abundances). ~$^a$Absorption \\column in units of $10^{21}$ cm$^{-2}$.  ~$^b${\it Observed} fluxes in the 0.3--10\,keV band, in units of $10^{-13}$ erg s$^{-1}$ cm$^{-2}$. ~$^c${\it Observed} and \\{\it unabsorbed} luminosities in the 0.3--10\,keV band, in units of $10^{41} \ergsec$ (assuming a distance of 19.8\,Mpc). ~$^{\dag}$$N_H$ component fixed. \\~$^{\ddag}$$N_H$ components fixed at $10^{21}$ cm$^{-2}$ and $10^{22}$ cm$^{-2}$ for {\it wabs} and {\it pcfabs} components respectively.  Errors correspond to 90 percent \\confidence limits for 1 parameter of interest. The best-fit model is shown in bold.\\
\end{tabular}
\label{table:n3310}
\end{table*}

The discrete source in the \xmmn data (named XMM-1) at $10^h27^m55.3^s$, $-43\deg54\arcm46\arcs$ is coincident with source 14 in the \chandra data \citep{lira02}. Since this source does not have sufficient counts for spectral fitting, we have computed its hardness ratios using the PN counts in three different energy bands using a similar method to that described in \citet{lira02} in order to compare its properties in both observations. We compute $log$(0.3--1\,keV)/(1--2\,keV)=-0.26 and $log$(2--7\,keV)/(1--2\,keV)=-0.32, which, in terms of the \xmmn instrument response, is equivalent to a photon index of $\Gamma\sim$ 2.5 and an absorbing hydrogen column of $N_H\sim5\times10^{21}$ cm$^{-2}$.  This is softer than source 14 in the \chandra data, which shows $\Gamma\sim$ 2 (see Figure 4 in \citealt{lira02}), although the absorbing columns are similarly high.

If we assume a power-law photon index of $\Gamma$=2.5 and adopt an absorbing column of $N_H\sim5\times10^{21}$ cm$^{-2}$ (as measured from the both the \xmmn and \chandra hardness ratios), we get an observed flux in the 0.5--10\,keV range of $F_X\sim2.5\times10^{-14}$ erg s$^{-1}$ cm$^{-2}$ and an unabsorbed luminosity of $L_X\sim7\times10^{39} \ergsec$, within $\sim$ 20 percent of the \chandra luminosity for the same parameters. This luminosity places this source into the ULX regime, although this value is highly dependent on the large absorption correction estimated from the hardness ratios alone. 

\citet{lira02} speculate that the luminous X-ray sources in NGC~3256 are likely to be high-mass XRBs (HMXBs). They detect no short-term variability in the \chandra data expected from an accreting source, although variability on short timescales is not generally detected in \chandra observations of ULXs (c.f. \citealt{roberts04}). The position is coincident with an outer spiral arm of the galaxy and offset by $\sim$4 arcseconds from an apparent foreground star, but given that there is no evidence of either significant long- or short-term variability, this source could be an obscured background AGN. Using the log($N$)--log($S$) relation of \citet{hasinger01}, we calculate that in an annular region of $\sim$ 1.2 arcminute radius centred on NGC~3256, excluding the 0.45 arcminute radius of the galaxy itself, we expect to detect 0.02 background AGN at the 2--10\,keV flux of this source ($F_X\sim1.6\times10^{-14}$ erg s$^{-1}$ cm$^{-2}$). This suggests that the source is likely to be associated with NGC~3256.

\subsection{NGC~3310}
\label{sec:n3310_results}

Figure~\ref{fig:n3310} [top] shows the soft (0.3--2\,keV, left) and hard (2--6\,keV, right) \xmmn EPIC images of NGC~3310.  The majority of the X-ray emission comes from the bright central region of the galaxy, corresponding to the nucleus, starburst ring, inner disk, and outer spiral arm. There are also two symmetrically-positioned discrete sources of X-ray emission to the northwest and southeast of the nucleus (discussed below), which we denote as X-2 (south) following \citet{roberts00}, and X-3 (north) (although X-3 was not included in that paper due to an error in the original data analysis). X-3 was also detected in the \citet{zezas98} \rosat study (J103843.2+533107), although X-2 was not. 

A recent $\sim$30\,ks \chandra observation of NGC~3310 (Zezas et al. {\it in preparation}) has revealed the presence of 24 discrete sources with X-ray luminosities ranging from $10^{37}$--$10^{40} \ergsec$, most of which rather spectacularly trace the central starburst ring (Figure~\ref{fig:n3310} [lower left]). There is also evidence in the \chandra images of extended diffuse emission. Figure~\ref{fig:n3310} [lower right] shows the \chandra X-ray contours overlaid on an \xmmn OM UV image of NGC~3310 showing again, as in the case of NGC~3256, that the UV and X-ray emission are well correlated. 

We have fitted the MOS and PN spectra for the main body of the galaxy (excluding the two discrete sources) following the same fitting procedures for the \xmmn spectra as used for NGC~3256, albeit only in the 0.3--6\,keV range due to the high level of background flaring in this observation (see section~\ref{sec:analysis}). The results are shown in Table~\ref{table:n3310}, with fluxes and luminosities extrapolated to the 0.3--10\,keV band. For this galaxy we have also fitted the $N_H$ column associated with the coolest MEKAL component, as leaving this parameter free does not result in unrealistic sub-Galactic values as was the case for NGC~3256. 

To begin with, we fitted the data using models where all $N_H$ columns were left as free parameters. This method produced very good fits to the data with both M+PL ($\chi^2_{\nu}$=0.89) and M+M+PL ($\chi^2_{\nu}$=0.87) models, with a statistical improvement at the 98.2 percent level for the addition of the second thermal component. However, the fitted power-law component had very little intrinsic column (3$\times10^{20}$ cm$^{-2}$) compared to the thermal components (5$\times10^{21}$ and 4$\times10^{22}$ cm$^{-2}$ for the cool and warm components respectively). \chandra observations of starforming galaxies have shown that point sources typically have non-negligible absorption e.g. the NGC~3256 discrete sources have columns in the range $10^{21}$ to $10^{22}$ cm$^{-2}$ (see Table~3, \citealt{lira02}). Additionally, if we assume that the point sources are related to the same stellar population as the thermal components, we would not expect to see such large differences in their associated column densities. These models also implied that at energies $<$ 0.5\,keV, the emission was dominated by the power-law component from point sources, but this is inconsistent with soft (0.3--2\,keV) and hard (2--8\,keV) \chandra images, which show a large contribution from apparently diffuse emission at soft energies. 

In order to achieve a more physically realistic fit, we re-fitted the data with models where the power-law absorption component was fixed. Firstly, we fixed it to the value obtained from the simple power-law plus absorption model ($N_H$=1.19$\times10^{21}$ cm$^{-2}$), and fitted the remainder of the spectrum with one and two thermal components (models 1 \& 2 in Table~\ref{table:n3310}). Both models gave good fits to the data, with a $>$ 99 percent improvement with the addition of the second thermal component. The absorption values obtained with these fits are much more realistic than our initial fits, with $N_H$ values in the range of 0.4--3$\times10^{21}$ cm$^{-2}$.

Secondly we tried a model in which the power-law absorption comprised one $N_H$ component fixed at 1$\times10^{21}$ cm$^{-2}$, and a partial covering absorption component fixed at 1$\times10^{22}$ cm$^{-2}$ with a covering fraction of 0.5, i.e. effectively some power-law emission experiencing moderate columns and some experiencing higher absorption. This resulted in improved fits (models 3 \& 4 in Table~\ref{table:n3310}). Again, the absorption values measured with these fits are reasonable with $N_H\sim6\times10^{20}$ for the cool MEKAL component and $N_H\sim2\times10^{21}$ for the warm MEKAL component. We therefore adopt this M+M+PL model (model 4, shown in bold in Table~\ref{table:n3310}) as the best-fit model for this data. In this case, if we consider more realistic model errors i.e. 90\% confidence limits for 2 parameters of interest ($\Delta\chi^{2}$=4.61), the absorption and temperature parameters for the thermal components are slightly less well constrained than those of NGC~3256. While we can only gain an upper limit to the $N_H$ component tied to the warm thermal component ($N_H\la4\times10^{21}$ cm$^{-2}$), the constraints on the temperature remain unchanged. The parameters for the cool component are only slightly less well constrained with $N_H=0.3-3.3\times10^{21}$ cm$^{-2}$ and $kT$=0.14--0.30\,keV.

\begin{figure}
\centering
\scalebox{0.34}{\rotatebox{270}{\includegraphics{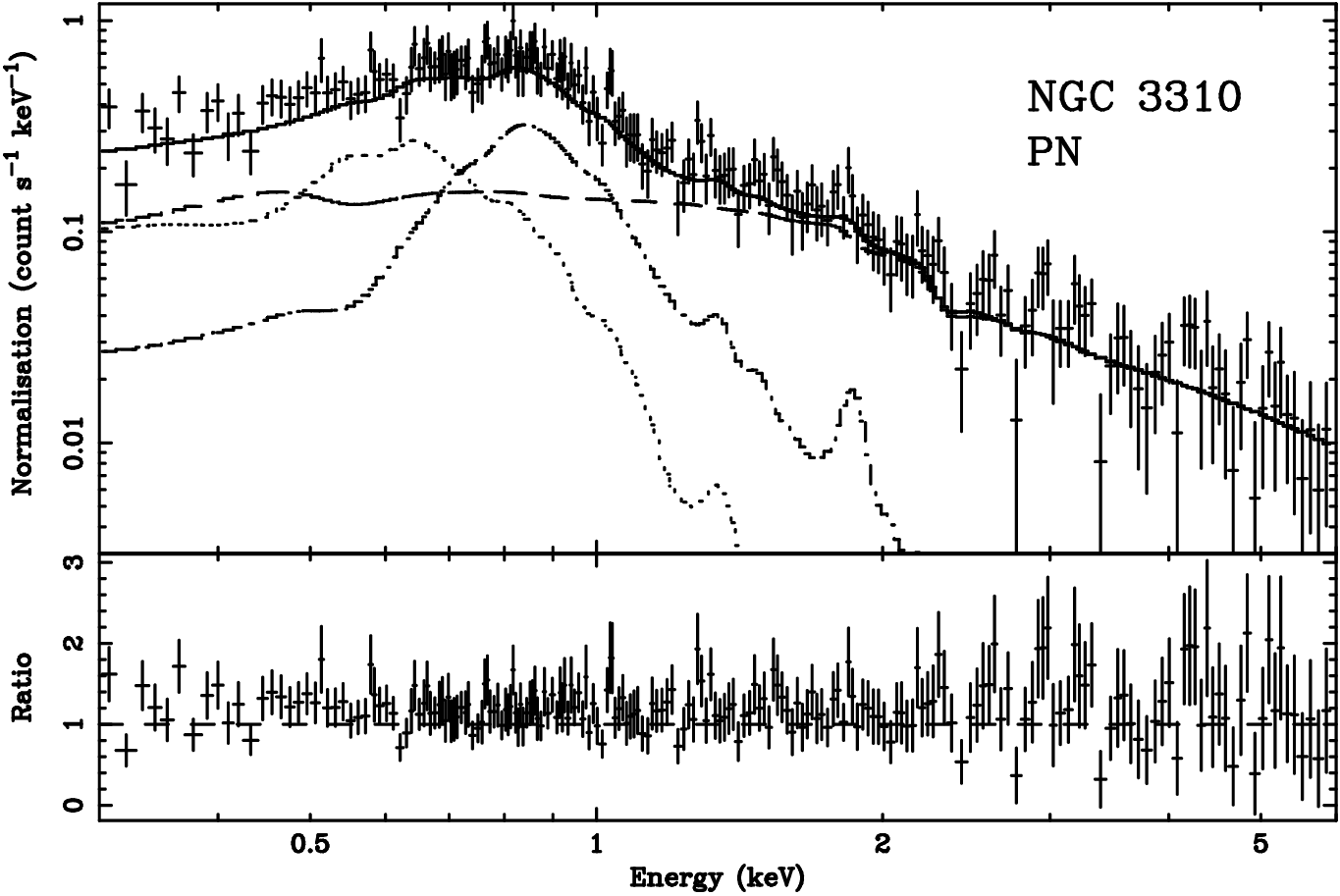}}}
\scalebox{0.34}{\rotatebox{270}{\includegraphics{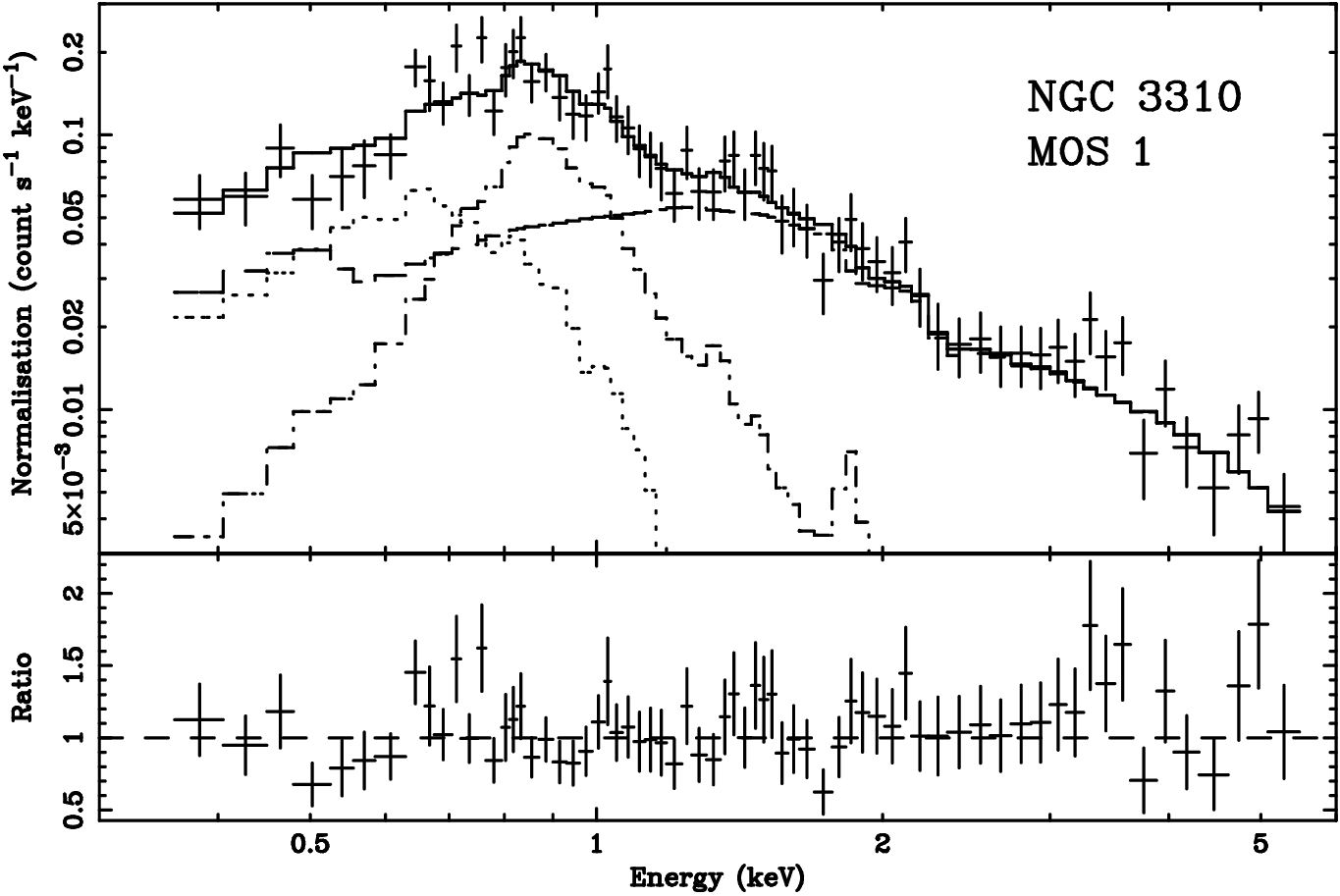}}}
\scalebox{0.34}{\rotatebox{270}{\includegraphics{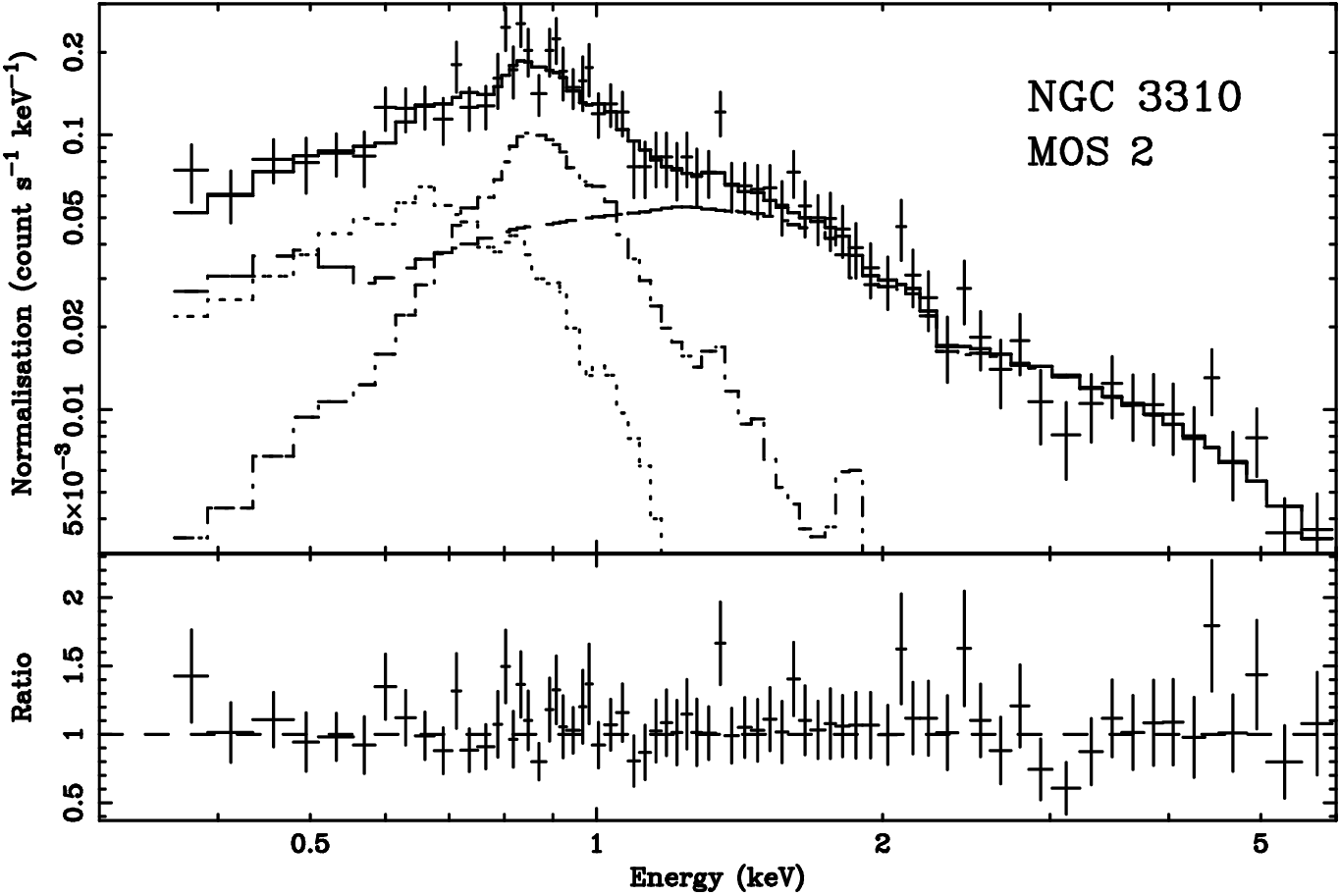}}}
\caption{\xmmn spectra and folded best-fit model (M+M+PL) of NGC~3310: PN [top], MOS 1 [middle], MOS 2 [bottom]. The model components are denoted by dashed (power-law), dotted (cool MEKAL) \& dot--dashed (warm MEKAL) lines.}
\label{fig:3310spec}
\end{figure}

The PN and MOS spectra are plotted with the best-fit model in Figure~\ref{fig:3310spec}, and Table~\ref{table:n3310_comps} lists the observed fluxes, unabsorbed luminosities and overall percentage contributions to the total of each from the three spectral components. The power-law continuum is the dominant component over the 0.3--10\,keV range, contributing $\sim$80 percent of the total observed flux and unabsorbed luminosity of the galaxy. We note that we were unable to test for the presence of Fe-K lines after excluding data above 6\,keV.

\begin{table}
\caption{Fluxes, luminosities and percentage of total emission from the three components of the best-fit M+M+PL model of NGC~3310.}
 \centering
  \begin{tabular}{lccccc}
\hline
Component  &$F_X$  &\% of total & $L_X$ & \% of total \\
\hline
Cool MEKAL & 1.50  & 9          & 1.09  & 9           \\
Warm MEKAL & 2.00  & 11         & 1.98  & 16          \\
Power-Law   & 13.80 & 80         & 9.48  & 75          \\
\hline
Total      & 17.30 & 100        & 12.55 & 100         \\
\hline
\end{tabular}
\begin{tabular*}{7.5cm}{@{}l}
Notes:  {\it Observed} fluxes in the 0.3--10\,keV band, in units of \\$10^{-13}$ erg s$^{-1}$ cm$^{-2}$. {\it Unabsorbed} luminosities in the \\0.3--10\,keV band, in units of $10^{40} \ergsec$ (assuming a \\ distance of 19.8\,Mpc).\\
\end{tabular*}
\label{table:n3310_comps}
\end{table}

\subsubsection{Comparison with previous work}

In the \rosat and \asca observations of NGC~3310 reported by \citet{zezas98}, the combined spectrum of the integrated emission from NGC~3310 was well-fit with either a double RS thermal plasma ($kT$=0.80/14.98\,keV) or one comprised of a power-law plus RS thermal plasma ($\Gamma$=1.44, $kT$=0.81\,keV), with both components affected by a single absorbing column. These parameters are partly consistent with the \xmmn results; the power-law slope is slightly harder and the thermal plasma component similar (to within the 90 percent confidence limits) to the warm MEKAL component in our best-fit M+M+PL model. However, the main difference is that whereas they fitted a hot thermal plasma plus a low absorbing hydrogen column ($N_H\sim0.17\times10^{21}$ cm$^{-2}$), the improved quality of the data in the \xmmn observation at the soft energies means that we are able instead to model the soft emission with a cool thermal plasma with a higher absorbing column. There is also a marked difference in the measured fluxes and luminosities in \rosat/\asca and \xmmn datasets. Even if we consider the nearly identical two-component PL+M/PL+RS models, there is a $\sim$ 40 percent reduction in the {\it observed} flux in the \xmmn data compared with the \rosat/\asca analysis ($\sim3\times10^{-12}$ erg s$^{-1}$ cm$^{-2}$). It is likely that this is due to the difference in size of the extraction regions used. The \rosat spectra were extracted using a 1.5 arcminute radius region and the \asca regions were larger at 2.7 and 5.5 arcminutes radius for the SIS and GIS respectively \citep{zezas98}, whereas we have used a 40 arcsecond radius circular source region to avoid contamination from the two discrete sources. 

In order to make a direct comparison of the observed fluxes, we have attempted to replicate the \asca/\rosat results by extracting \xmmn spectra in an aperture with the same radius as the largest \asca aperture (5.5 arcminutes) together with as large an annular background region as possible given the size of the MOS and PN chip arrays. However, this turns out to be impractical, as the \xmmn spectra are dominated by the high background flux (especially at energies $>2$\,keV), resulting in low-quality background-subtracted data which cannot constrain the spectral shape, and hence cannot constrain the flux from a larger region.

\subsubsection{Discrete Source Properties}

The two discrete sources to the south (X-2) \& north (X-3) of the main body of the galaxy have insufficient counts for detailed spectral fitting ($\sim$130 and $\sim$110 in each MOS camera respectively). However, to estimate their spectral shapes, we have co-added and fitted the MOS spectra (X-3 is located on a chip gap in the PN) with simple power-law plus Galactic absorption models. Their \xmmn positions, power-law photon indices, observed fluxes and unabsorbed luminosities are shown in Table~\ref{table:n3310_ulx}. The X-ray luminosities of these sources are high ($7\times10^{39} \ergsec$ \& $5\times10^{39} \ergsec$, 0.3--10\,keV), putting them well into the ULX regime. Although not shown in the OM image (Figure~\ref{fig:n3310} [lower right]), they are both coincident with regions of UV emission, which is indicative of star formation activity though there are no obvious UV counterparts in this data.  Although this \xmmn observation is very flare-contaminated, we tested the variability of each source in the \xmmn data by deriving short-term light curves. The data from the MOS cameras were co-added to improve the signal-to-noise ratio, and the data were tailored so that each bin had at least 20 counts after background subtraction. We performed $\chi^2$ tests to search for large amplitude variability against the hypothesis of a constant count rate, but neither source showed any significant variability ($\ga 3\sigma$). In order to search for longer-term variability, we have made a direct comparison between their observed fluxes in this \xmmn observation and the 1995 \rosat HRI 41\,ks observation, both in the 0.3--2.4\,keV band as this is the energy range covered by both mission.  We derived fluxes for the \rosat HRI observation with {\small webPIMMS} using the observed count rate from \citet{roberts00} for X-2, and a new count rate estimate for X-3 from the archival HRI data, assuming the absorbed power-law continuum slope measured in the \xmmn observation and normalized to the 0.3--2.4\,keV band. During this $\sim$6 year period, the flux for X-3 has decreased by $\sim$57 percent, while that of X-2 has increased by $\sim$65 percent.  Variability on this timescale is consistent with that found for other ULX (c.f. \citealt{roberts04}), and supports the scenario that these sources are single black-hole XRBs rather than concentrated regions of star formation, as multiple sources such as a cluster of XRBs would not be expected to all change their output simultaneously.

\begin{table}
\caption{Discrete Sources in NGC~3310.}
 \centering
  \begin{tabular}{@{}lcccc@{}}
\hline
Source  &Position (J2000)                         &$\Gamma$              &$F_X$ &$L_X$ \\
\hline
X-2     &$10^h38^m50.2^s$ $+53\deg29\arcm25\arcs$ &1.49$^{+0.96}_{-0.72}$ &1.47   &7.00 \\
X-3     &$10^h38^m43.3^s$ $+53\deg31\arcm00\arcs$ &1.66$^{+0.88}_{-0.75}$ &1.05   &5.01 \\
\hline
\end{tabular}
\begin{tabular*}{8.4cm}{@{}l}
Notes:  {\it Observed} fluxes in the 0.3--10\,keV band, in units of $10^{-13}$ \\erg s$^{-1}$ cm$^{-2}$. {\it Unabsorbed} luminosities in the 0.3--10\,keV band, \\ in units of $10^{39} \ergsec$ (assuming a distance of 19.8\,Mpc).\\
\end{tabular*}
\label{table:n3310_ulx}
\end{table}

\section{Discussion}
\label{sec:discuss}

\begin{table*}
\caption{Comparison of the properties of NGC~3256, NGC~3310 and the Antennae (NGC~4038/9).}
 \centering
  \begin{tabular}{lccccccc}
\hline
Parameters                     && NGC~3256                 && NGC~3310            && Antennae (NGC~4038/9)   & Refs \\
\hline
Distance (Mpc, $H_0$=75\,km\,s$^{-1}$\,Mpc) && 35.4        && 19.8                && 19.3                    & 1,2\\ 
\\
Merger Type                    && Major merger - two       && Merger with dwarf   && Major merger - two      & 3,4,6\\
                               && equal mass galaxies      && companion           && equal mass galaxies     & \\ 
\\
Merger Age                     && $\sim5\times10^{8}$\,yr  && $\la10^{8}$\,yr     && $2-5\times10^{8}$\,yr  & 3,4,5,6\\
\\
Starburst Age                  && $5-25\times10^{6}$\,yr   && $10^{7}-10^{8}$\,yr && $\sim5-100\times10^{6}$\,yr & 3,5,7\\
\\
Dynamical Mass ($10^{10}M_{\odot}$) && $\sim5$             && $\sim2.2$           && $\sim8$                 & 3,8\\
\\
$L_{FIR}$ ($L_{\odot}$)$^{\dag}$    && $1.8\times10^{11}$       && $2.0\times10^{10}$  && $2.9\times10^{10}$      & 1\\
\\
SFR$_{IR}$ ($M_{\odot}$ yr$^{-1}$)$^{\ddag}$ && 32.6           && 3.6                 && 5.2                     & \\
\\
$L_X$$^*$ ($\ergsec$)          && 4.5$\times10^{41}$       && 1.3$\times10^{41}$  &&  6.7$\times10^{40}$     & \\
\\
M$_1$$^*$ ($kT$/$keV$)         && 0.6 (12\%)               && 0.3 (9\%)           &&  0.4 (17\%)             & \\
\\
M$_2$$^*$ ($kT$/$keV$)         && 0.9 (70\%)               && 0.6 (16\%)          &&  0.7 (14\%)             & \\
\\
$\Gamma$$^*$                   && 2.0 (18\%)               && 1.8 (75\%)          &&  1.6 (69\%)             & \\
\hline
\end{tabular}
\begin{tabular}{l}
References: (1) \citet{sanders03}; (2) \citealt{zezas02a}; (3) \citet{lipari00}; (4) \citet{balick81}; \\(5) \citet{whitmore99}; (6) \citet{barnes88}; (7) \citet{elmegreen02}; (8) \citet{kregel01}. ~$^{\dag}$FIR luminosities calculated \\using FIR fluxes derived from the $60\mu$ and $100\mu$ IR flux densities of \citet{sanders03} using the relation of \\\citet*{helou85}: $FIR=1.26\times10^{-11}(2.58S_{60\mu}+S_{100\mu}) \ergcms$.~$^{\ddag}$SFRs determined using the relation \\of \citet{kennicutt98}: SFR=${L_{FIR}/2.2\times10^{43}} M_{\odot}$\,yr$^{-1}$.~$^*${\it Unabsorbed} luminosities in the 0.3--10\,keV band and spectral parameters from \\the best-fit models. Percentage contributions from the components to the total unabsorbed luminosity are shown in parentheses.\\
\end{tabular}
\label{table:summary}
\end{table*}

Our results are consistent with other local SBGs studied with \xmmn and {\it Chandra}, e.g. NGC~253 (\citealt{pietsch01}; \citealt{strickland00b}; \citealt{strickland02}); M82, (\citealt{stevens03}; \citealt{matsumoto01}) and the closest merger system NGC~4038/4039 (The Antennae), (\citealt*{fabbiano01}; \citealt{fabbiano03}; \citealt{zezasfab02}; \citealt{zezas02a}; \citealt{zezas02b}). The X-ray emission from these systems are well-fitted with thermal components ranging between $\sim$0.2--0.9\,keV plus harder ($>$ 2\,keV) emission dominated by power-law continua from discrete sources.

Even with the relatively short exposures of these \xmmn observations, especially in the case of NGC~3310 where there was substantial background flare contamination in the data, we are able to improve the definition of the plasma temperatures and power-law slopes compared to earlier \rosat and \asca data for these galaxies. Additionally, the information from the spatially-resolved \chandra data has allowed us to deconvolve this spectral information and properly assess the contributions from point sources and diffuse components.

An interesting result of this study are the different temperatures of the thermal components in NGC~3256 and NGC~3310. To investigate the reasons for this, we have summarized the known properties of each system in Table~\ref{table:summary}. We have also included details for the Antennae system (NGC~4038/39), as this is the best example of a close merger system of two equal-mass galaxies. For the purposes of this comparison, we selected data from the \chandra archive (Obs ID 3042) to provide us with a deep observation ($\sim$ 70\,ks) unaffected by background flaring.  We extracted a cumulative spectrum from all X-ray sources in a central 1 arcminute diameter circular region using the {\small CIAO} tool {\small ACISSPEC} (which provides RMFs and ARFs weighted over the differing calibration across the extraction zone), and used adjacent source-free regions to provide a background measurement. We fitted the spectrum in the 0.3--10\,keV range with the same spectral models used for NGC~3256 and NGC~3310, and the data are well-fit with a power-law ($\Gamma$=1.6) and two thermal plasmas ($kT$=0.40/0.74\,keV), with no absorption required above the Galactic value ($3.4\times10^{20}$ cm$^{-2}$, \citealt{stark92}). The power-law component can be entirely accounted for by the systems' population of compact point sources \citep{zezas02a}, and the plasma temperatures are consistent with the detailed spatial and spectral analysis of the diffuse emission reported in \citet{fabbiano03}. 

On inspection, albeit on the basis of only three systems, there does appear to be a correlation between the type of merger, the age of the starburst component and the remaining properties ($L_{FIR}$, $L_X$ \& plasma temperature). NGC~3256 is a major merger with a young starburst, and displays high levels of FIR and X-ray emission together with hot plasma temperatures. Conversely, NGC~3310 is a minor merger with an older starburst component, has lower FIR and X-ray luminosities and displays cooler plasma temperatures. The Antennae system, on the other hand, has intermediate aged star-formation spread over $\sim5-100$\,Myr \citep{whitmore99}, and displays plasma temperatures between those of NGC~3256 and NGC~3310. However, although the Antennae is an equal-mass merger like NGC~3256, its FIR and X-ray luminosities are a factor of $\sim$ 6 lower; given that the Antennae has the greater mass of the two, this suggests a decrease in the power of the starburst with time that is independent of the total mass of the system. There is also a difference in the relative contributions from the spectral components to the total intrinsic X-ray luminosities of the systems. While NGC~3310 and the Antennae are dominated by the power-law component, the intrinsic emission in NGC~3256 is dominated by the hot thermal plasma due to its high absorption column.  The ratio of intrinsic X-ray and FIR luminosities is also different between the the galaxies; while NGC~3256 and the Antennae both have $L_{X}$/$L_{FIR}\sim6\times10^{-4}$, NGC~3310 has a greater relative X-ray luminosity with $L_{X}$/$L_{FIR}\sim2\times10^{-3}$.

These properties pose a number of interesting questions. For example, does the apparent correlation between plasma temperature and starburst age simply indicate that the hotter phases of the ISM are escaping the systems at earlier times after the onset of the starburst episodes? Or could the higher masses and hence deeper gravitional potential wells of NGC~3256 and the Antennae mean that the hotter components are retained in these systems for longer?  Is the highly absorbed hot thermal plasma in NGC~3256 also a consequence of the younger age of the starburst i.e. has insufficient time elapsed for the neutral absorbing material to be blown away from the dense central regions? Larger samples of high-quality X-ray data from merging galaxies would be required to investigate these trends properly.

\subsection{Thermal X-ray emission}
\label{sec:discuss_therm}

The soft emission in SBGs is thought to arise primarily in hot gas from both the interaction of supernova-driven galactic winds with the ambient ISM and from SNRs themselves. Thermal components are also seen in more distant examples of merger systems, e.g. \citet{franceschini03} have recently reported the presence of soft thermal plasma emission in all galaxies in a survey of ten ULIRGs. 

The soft emission in both NGC~3256 and NGC~3310 is well-modelled by multi-temperature thermal plasma models. Both galaxies show a trend of increasing absorbing column densities with increasing plasma temperature, similar to that found in other SBGs.  Although we cannot spatially resolve the different emitting regions with the \xmmn data, \citet{lira02} were able to show that in the \chandra data, the hot 0.9\,keV component dominates in the inner $\sim$2\,kpc of NGC~3256, while the cooler 0.6\,keV component dominates at radii $\ga$ 2.5\,kpc (adjusted to a distance of 35.4\,Mpc). This evidence implies that the absorbed hot phase of the ISM arises from supernova-heated gas possibly within the galactic disk, while the cooler component with less absorption may arise in outflowing galactic winds. The thermal components in NGC~3310 have characteristics consistent with this picture, although spatially resolved spectroscopy are required to pin down the different emitting regions. 

As this study demonstrates, particularly in the case of NGC~3310 where we have successfully modelled a two-temperature thermal plasma for the first time, the increased photon statistics and hence spectral quality we can acquire with the current generation of X-ray telescopes is revealing increasingly complex temperature profiles of the thermal emitting gas. However, we must bear in mind that this type of modelling with {\it distinct} thermal plasma temperatures is likely to be over-simplification of the real state of the hot gas in these systems; instead a {\it phase continuum of states} with temperatures between $T\sim10^5-10^8$\,K is more likely to exist \citep{strickland00a}.

\subsection{Hard X-ray emission}
\label{sec:discuss_hard}

The \chandra data for these galaxies has resolved the majority of the hard power-law flux in both of these galaxies into a population of luminous compact sources embedded in the diffuse emission (a full analysis of the \chandra data will be presented in Zezas et al. {\it in preparation}).  This agrees with other recent studies of SBGs, where large numbers of ULXs have been found to be associated with regions of current star formation in e.g. The Cartwheel \citep{gao03}; The Antennae (NGC~4038/39; \citealt{zezas02b}); NGC 4485/90 \citep{roberts02}; Arp299 (NGC~3690; \citealt*{zezas03}), suggesting that they are associated with young stellar populations and are therefore likely to be HMXBs going through a phase of thermal-timescale mass transfer \citep{king04}. Such studies serve to emphasize the dominant contribution of a small number ($\sim$10) of probable black hole XRBs to the flux of the power-law continuum component in the most luminous SBGs.  We note that this disagrees with the suggested model of \citet{persic02}, where the power-law continuum should originate in a large number of neutron star binaries ($>$ 500 to reach $\sim10^{41} \ergsec$ if they are Eddington-limited). Future composite models for SBGs should therefore consider this subtle change in the origin of the power-law component.

Although these few sources dominate the emission at high energies, there is also likely to be a contribution from hotter gas that we cannot detect with this data. In the case of NGC~3256, \citet{lira02} have shown that after point source subtraction, the hard component is best modelled with a $\sim$4\,keV thermal plasma, though there is likely to be a contribution from unresolved point sources. Similary in NGC~253, emission lines from thermal plasma component $>$ 5\,keV are detected in the \xmmn RGS (Reflection Grating Spectrometer) and a hot plasma of $\sim$6\,keV is required in the EPIC spectra of the nuclear region \citep{pietsch01}.

\subsection{Starburst/AGN activity}
\label{sec:discuss_AGN}

We find no strong evidence for the presence of an AGN in either NGC~3256 or NGC~3310. Hidden AGN activity can be detected at X-ray energies by the presence of a highly absorbed ($N_H>10^{22}$ cm$^{-2}$) hard component and/or a strong neutral Fe-K line at $\sim$6.4\,keV originating in the nucleus. In the \xmmn ULIRG survey of \citet{franceschini03}, roughly half of the systems studied showed evidence of AGN activity, confirming their composite nature but suggesting that the starburst component is dominant as the thermal signatures were detected in all systems. Evidence for AGN activity has also recently been found in nearby ULIRGs with the detection of strong neutral Fe-K emission lines e.g. NGC~6240 (\citealt{boller03}; \citealt{komossa03}) and NGC~3690 \citep{ballo04}. 

However, the power-law components in the best-fit models of both galaxies do not require such high absorbing columns (NGC~3256, $N_H\sim1.5\times10^{21}$ cm$^{-2}$; NGC~3310, $N_H\sim1.2\times10^{21}$ cm$^{-2}$), and have been shown to originate in the discrete source populations. There {\it is} marginal evidence of Fe-K line emission in NGC~3256, with an unconstrained $\sim$6.5\,keV line energy and a 90 percent upper limit of $\sim$1\,keV.  This is consistent with the equivalent widths observed in some Compton-thick type 2 AGN, which can be as high as $>$1\,keV (e.g. \citealt*{molendi03}). However, hot thermal starburst components can also give rise to helium-like $\sim$6.7\,keV emission from  type Ib/IIa SNRs (e.g. \citealt{behar01}). This type of emission is seen in M82 \citep{stevens03} and the starburst compent of NGC~4945 \citep*{schurch02}.  However, this is only a weak upper limit on the equivalent width of a 1$\sigma$ line detection, and as we are unable to distinguish between the two types of line emission in these data, the question of the presence of a hidden AGN remains open.

\section{Conclusions}
\label{sec:conclusions}

In this paper, we have studied the X-ray spectral properties of the two nearby starburst merger galaxies NGC~3256 and NGC~3310.  We have used the broad-band (0.3--10\,keV) spectra from \xmmn together with complementary high-spatial-resolution imaging data from \chandra to assess the various stellar and non-stellar contributions to the X-ray emission. We have shown that the emission $\la$ 2\,keV is dominated by thermal emission believed to arise in the interaction of galactic winds with the ISM and SNRs, which can be well-modelled by thermal plasma models with representative temperatures of $\sim$0.3--0.9\,keV.  The emission between 2--10\,keV is dominated by power-law emission from the bright ULX population, with the remainder likely to come from fainter unresolved point sources and a hotter plasma.  These results are consistent with the models of \citet{persic02}, albeit with the slight modification that ULXs dominate the high energy emission, and provide yet further evidence of the complex nature of SBGs.  The more thorough understanding of local merger and starbursting systems obtained in studies such as this will certainly assist in our interpretation of the very photon-limited X-ray emission we currently detect from high-redshift merger systems.

\section*{Acknowledgments}

We thank the referee David Strickland for constructive comments which have helped improve this paper.  This work is based on observations obtained with {\it XMM-Newton}, an ESA science mission with instruments and contributions directly funded be ESA and NASA. LPJ acknowledges the support of a PPARC studentship.

\label{lastpage}

\bibliographystyle{mn2e}
{}

\end{document}